\documentclass{article}

\usepackage{authblk}
\usepackage{xcolor}
\usepackage{url}
\usepackage{graphicx}
\usepackage[numbers,sort]{natbib}
\usepackage{enumitem}
\usepackage{amssymb}
\usepackage{amsmath}
\usepackage{algorithm}
\usepackage{algpseudocode}
\usepackage{bm}
\usepackage{bbm}
\usepackage{float}
\usepackage{wrapfig}
\usepackage{booktabs}  
\usepackage{multirow}
\usepackage{listings}
\usepackage[T1]{fontenc}
\usepackage{microtype} 
\usepackage{hyperref} 

\ifdefined\ShowNotes
  \newcommand{\colornote}[3]{{\color{#1}\bf{#2 #3}\normalfont}}
\else
  \newcommand{\colornote}[3]{}
\fi

\definecolor{darkred}{rgb}{0.7,0.1,0.1}
\definecolor{darkgreen}{rgb}{0.1,0.5,0.1}
\definecolor{cyan}{rgb}{0.7,0.0,0.7}
\definecolor{dblue}{rgb}{0.2,0.2,0.8}
\definecolor{maroon}{rgb}{0.76,.13,.28}
\definecolor{burntorange}{rgb}{0.81,.33,0}
\definecolor{royalpurple}{rgb}{0.47,.31,0.66}

\ifdefined\ShowNotes
  
\else
  
\fi

\definecolor{dkgreen}{rgb}{0,0.6,0}
\definecolor{gray}{rgb}{0.5,0.5,0.5}
\definecolor{light-gray}{gray}{0.95}
\definecolor{mauve}{rgb}{0.58,0,0.82}
\definecolor{backcolour}{rgb}{0.95,0.95,0.92}

\lstdefinelanguage{CUDACPP}{
  language=C++,
  morekeywords={__global__, __host__, __device__, __shared__, blockIdx, blockDim, threadIdx, gridDim},
  morecomment=[l][\color{magenta}]{\#},
}

\lstdefinestyle{pythonstyle}{
  language=Python,
  frame=tb,
  aboveskip=3mm,
  belowskip=3mm,
  showstringspaces=false,
  columns=flexible,
  basicstyle={\ttfamily\footnotesize},
  numbers=none,
  numberstyle=\footnotesize\color{gray},
  keywordstyle=\color[rgb]{0.13,0.29,0.53},
  commentstyle=\color[rgb]{0.13,0.55,0.13},
  stringstyle=\color[rgb]{0.31,0.60,0.02},
  breaklines=true,
  breakatwhitespace=true,
  tabsize=4
}

\definecolor{tabblue}{HTML}{4e79a7}
\definecolor{tabred}{HTML}{e15759}
\hypersetup{
    colorlinks=true,
    linkcolor=tabred,
    citecolor=tabblue
}

\title{\textsc{Mixture-of-Kittens}: MoE Megakernel for NVL72s}

\author[1,2]{Stuart H. Sul}
\author[1,2]{Nash Brown}
\author[2]{Henry Wildermuth}
\author[2]{William Lin}
\author[2]{\authorcr Federico Cassano}
\author[1]{Christopher Ré}
\affil[1]{Department of Computer Science, Stanford University}
\affil[2]{Cursor Research}

\begin{document}

\maketitle

\begingroup
\renewcommand{\thefootnote}{}
\footnotetext{Correspondence to: \texttt{ssul@cs.stanford.edu}.}
\endgroup

\begin{abstract}

AI accelerator systems are rapidly consolidating into scale-up architectures, where tens to thousands of GPUs communicate over high-bandwidth, single-hop fabrics.
%
We find that existing Mixture-of-Experts (MoE) training systems, optimized for conventional scale-out networks, transfer poorly to this setting, often running slower than a naive baseline built with PyTorch and NCCL.
%
With industry roadmaps pointing toward even larger scale-up domains, understanding the performance tradeoffs of this hardware regime is increasingly important.
%
We present Mixture-of-Kittens (MoK), an MoE training system designed for Nvidia NVL72.
%
MoK builds on three insights that unlock performance on scale-up domains: (1) choosing push- or pull-based communication per operator, (2) restructuring the computation-communication overlap, and (3) fully eliminating CPU-GPU synchronization.
MoK distills these insights into a single deterministic training megakernel that fuses token dispatch, shared and routed expert FFNs, and token combine.
%
Across MoE layer shapes from four widely used open-weight models, MoK delivers up to \(2.37\times\) the throughput of the strongest publicly available baseline. In a production run on 512 GPUs spanning multiple GB300 NVL72 racks, MoK improves end-to-end training throughput by \(1.41\times\).

\end{abstract}

\section{Introduction}
\label{sec:introduction}

\begin{figure}[t]
    \centering
    \vspace{-10pt}
    \includegraphics[
        width=0.9\textwidth
    ]{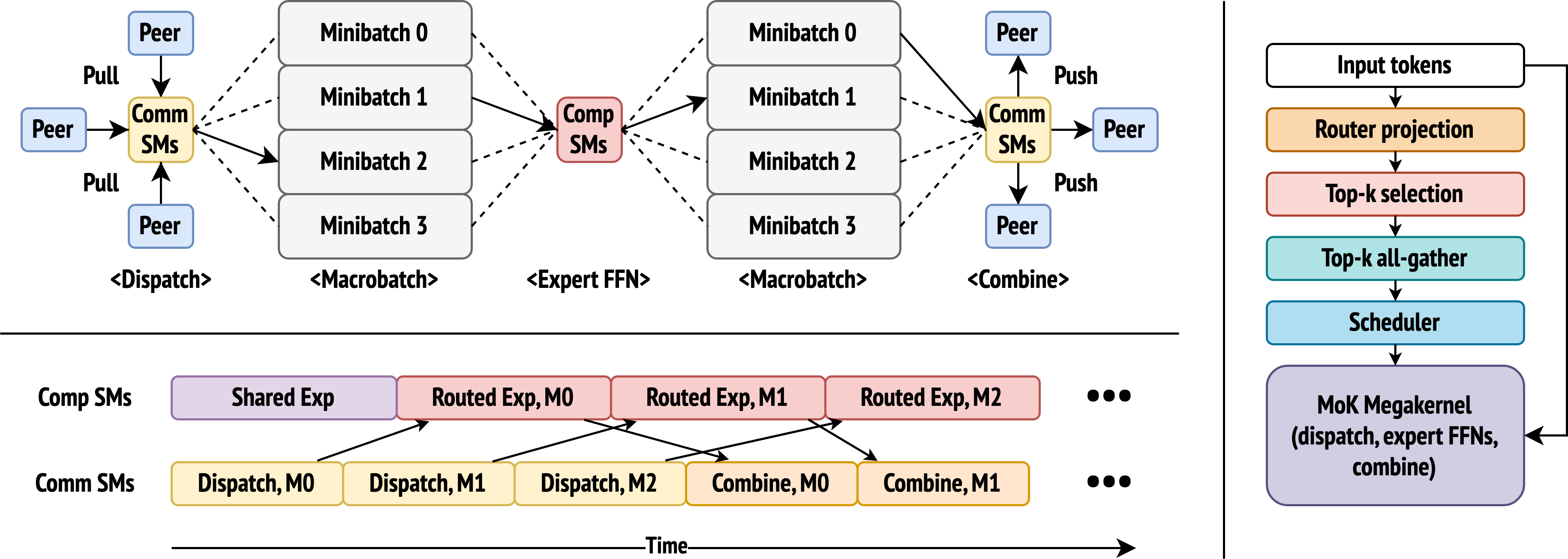}
    \vspace{-5pt}
    \caption{Overview of MoK. Top left: communication SMs pull tokens from peer GPUs into a fixed-size macrobatch ring buffer, expert FFNs process each minibatch, and communication SMs push the outputs back. Bottom left: MoK partitions SMs between computation and communication, and pipelines minibatches (M0, M1, \dots) to overlap expert computation with dispatch and combine. Right: before the megakernel launch, MoK performs GPU-side router projection, top-\(k\) selection and all-gather, and dispatch scheduling.}
    \label{fig:pull-figure}
    \vspace{-5pt}
\end{figure}

The hardware for training large language models (LLMs) is rapidly shifting toward scale-up architectures. Rather than having small groups of accelerators with a commodity scale-out network, vendors now place tens to thousands of accelerators inside a single high-bandwidth scale-up domain: examples include Nvidia's GB200/GB300 NVL72 \cite{nvidia2024gb200nvl72,nvidia2025gb300nvl72}, AMD's MI455X-based Helios \cite{amd2026helios}, Google's TPU Pod \cite{jouppi2023tpuv4}, and AWS Trn2 UltraServers \cite{aws2024trn2}. In these platforms, the interconnect stops behaving like a network and starts behaving like one large memory system, opening new opportunities for AI systems design while invalidating assumptions that existing designs were built on.

We present a study of this opportunity, focusing on Mixture-of-Experts (MoE) \cite{shazeer2017moe} training on Nvidia NVL72 systems \cite{nvidia2025gb300nvl72}. MoE has become a core component of frontier AI models and has been shown to significantly improve their capabilities \cite{kimi2026k27code,glm5team2026glm5,qwen2026qwen35,deepseekai2026v4}. However, when scaled across multiple GPUs, MoE becomes both communication- and compute-intensive, consequently consuming more than half of end-to-end execution time during large-scale training as reported by previous work \cite{pan2025fsmoe,yan2026megatroncore}. Meanwhile, Nvidia's roadmap points toward progressively larger scale-up domains, including NVL144, 576, and 1152 \cite{bhargava2026verarubinpod}. As the first generation in this family, NVL72 provides a timely platform for studying system tradeoffs that will become increasingly important in future hardware.

Prior work often accelerates distributed MoE by overlapping token communication with expert computation \cite{zhao2025deepep,zhang2025comet,yu2026hybridep}. However, we find that existing approaches either do not support scale-up platforms or become suboptimal on them. In nearly half of the configurations we evaluate, a naive baseline built from PyTorch and NCCL outperforms every evaluated alternative (Section~\ref{sec:moe-layer-benchmarks}). These results suggest that current MoE system designs do not transfer cleanly to the cost structure of scale-up hardware.

To understand this gap, we analyze MoE performance characteristics on NVL72s and distill our findings into three key insights that unlock performance on scale-up architectures:

\begin{enumerate}[itemsep=0.0pt, topsep=0pt]
    \item \textbf{Choosing the right communication direction} (Section~\ref{sec:choosing-the-right-communication-direction}). Unlike scale-out domains, scale-up fabrics make pull-based communication as efficient as push-based. This allows us to rethink each MoE operator's communication \textit{direction}, push or pull. We observe that prior designs make suboptimal direction choices that incur avoidable overheads in fine-grained communication, spending up to 18\% of its latency on completion signaling and 13\% of MoE runtime on pre-dispatch scheduling. By selecting the right set of directions (e.g., pull dispatch and push combine), we can keep signaling overhead under 1\%, build the schedule with up to 2.2$\times$ less overhead, and share one schedule table across all operators.
    \item \textbf{Restructuring the computation-communication overlap} (Section~\ref{sec:restructuring-computation-communication-overlap}). The scale-up domain enables computation-communication overlap across all 72 GPUs at granularities that were impractical over InfiniBand or RoCE. However, if the granularity is too fine, the tensor cores never saturate, while too coarse a granularity leaves dispatch and combine latency exposed. We find that the throughput-maximizing granularity ranges from 512 to 32,768 tokens across our workloads. By making granularity a first-class tunable parameter, and designing the GPU-side schedule, overlapping scheme, and megakernel fusion around it, we achieve up to \(3.53\times\) the throughput of a design fixed to a suboptimal granularity.
    \item \textbf{Fully eliminating CPU-GPU synchronization} (Section~\ref{sec:eliminating-cpu-gpu-synchronization}). The integrated CPUs on NVL72s run slower than those on DGX-based machines, with host-side execution of common PyTorch operations up to \(2.97\times\) slower. It is thus critical to remove CPU-GPU synchronization by moving memory buffer management entirely to the device side. To enable this, we cycle tokens through a fixed-size ring buffer with overlapped writes and reads, and integrate a ring-aware forward replay into the backward pass to recover overwritten tokens. Compared to over-allocation, our ring buffering adds only 1.7\% latency on average across our workloads.
\end{enumerate}

Based on these insights, we introduce \textbf{Mixture-of-Kittens (MoK)}, an MoE training system built from first principles for NVL72 platforms. MoK fuses all MoE computation and communication (i.e., dispatch, expert GEMMs, gated activation, and combine) into a single persistent megakernel \cite{spector2025megakernel,cheng2026mpk}. Beyond demonstrating the above three insights, MoK is a fully production-ready system. It provides features critical for real-world MoE training, including determinism, RDMA overlap for FSDP, MXFP8 support with fused quantization, fused router weight gradient computation, and tunable SM partitioning (Section~\ref{sec:additional-features-and-optimizations}).

We evaluate MoK across the shapes of four widely used open-weight models: Kimi K2.7, GLM 5.2, Qwen 3.5-397B-A17B, and DeepSeek V4 Pro. Against publicly available implementations that support NVL72s, MoK outperforms the strongest baseline by up to 2.37$\times$ (MXFP8 forward), 1.78$\times$ (MXFP8 backward), 1.92$\times$ (BF16 forward), and 1.58$\times$ (BF16 backward). In Cursor's production training stack on 512 GPUs across multiple GB300 NVL72 racks, MoK improves end-to-end training throughput by 1.41$\times$ in tokens per second per GPU over the previous DeepEP-based implementation.

MoK is open-sourced\footnote{\url{https://github.com/cursor/mixture-of-kittens}} and powers the training of Composer \cite{chan2026composer2}, Cursor's agentic coding model, across tens of thousands of GPUs. MoK is integrated into Nvidia NeMo AutoModel as the MoE execution backend \cite{nvidia2025automodel}.

In summary, we make the following contributions:
\begin{itemize}[itemsep=0pt, topsep=0pt]
    \item We present MoK, an open-source MoE training system that fuses all MoE computation and communication into a single deterministic megakernel, with production features including MXFP8 support, RDMA overlap for FSDP, and fused router weight gradient computation (Section~\ref{sec:mixture-of-kittens}).
    \item We identify three insights for efficient fine-grained computation-communication overlap on scale-up platforms: per-operator communication direction, an overlap scheme supporting arbitrary granularity, and on-device ring buffering to eliminate CPU-GPU synchronization (Sections~\ref{sec:choosing-the-right-communication-direction}--\ref{sec:eliminating-cpu-gpu-synchronization}).
    \item We demonstrate that MoK achieves up to 2.37$\times$ speedup over the strongest publicly available baselines across four widely used open-weight model shapes (Section~\ref{sec:moe-layer-benchmarks}), as well as a 1.41$\times$ end-to-end throughput improvement in a production training stack (Section~\ref{sec:End-to-end-benchmarks}).
\end{itemize}

\section{Background}
\label{sec:background}

In this section, we review MoE and expert parallelism, GPUs and scale-up architectures, and related work.

\subsection{MoE and Expert Parallelism}
\label{sec:moe-and-expert-parallelism}

\begin{wrapfigure}{R}{0.4\textwidth}
    \centering
    \vspace{-30pt}
    \includegraphics[width=0.4\textwidth]{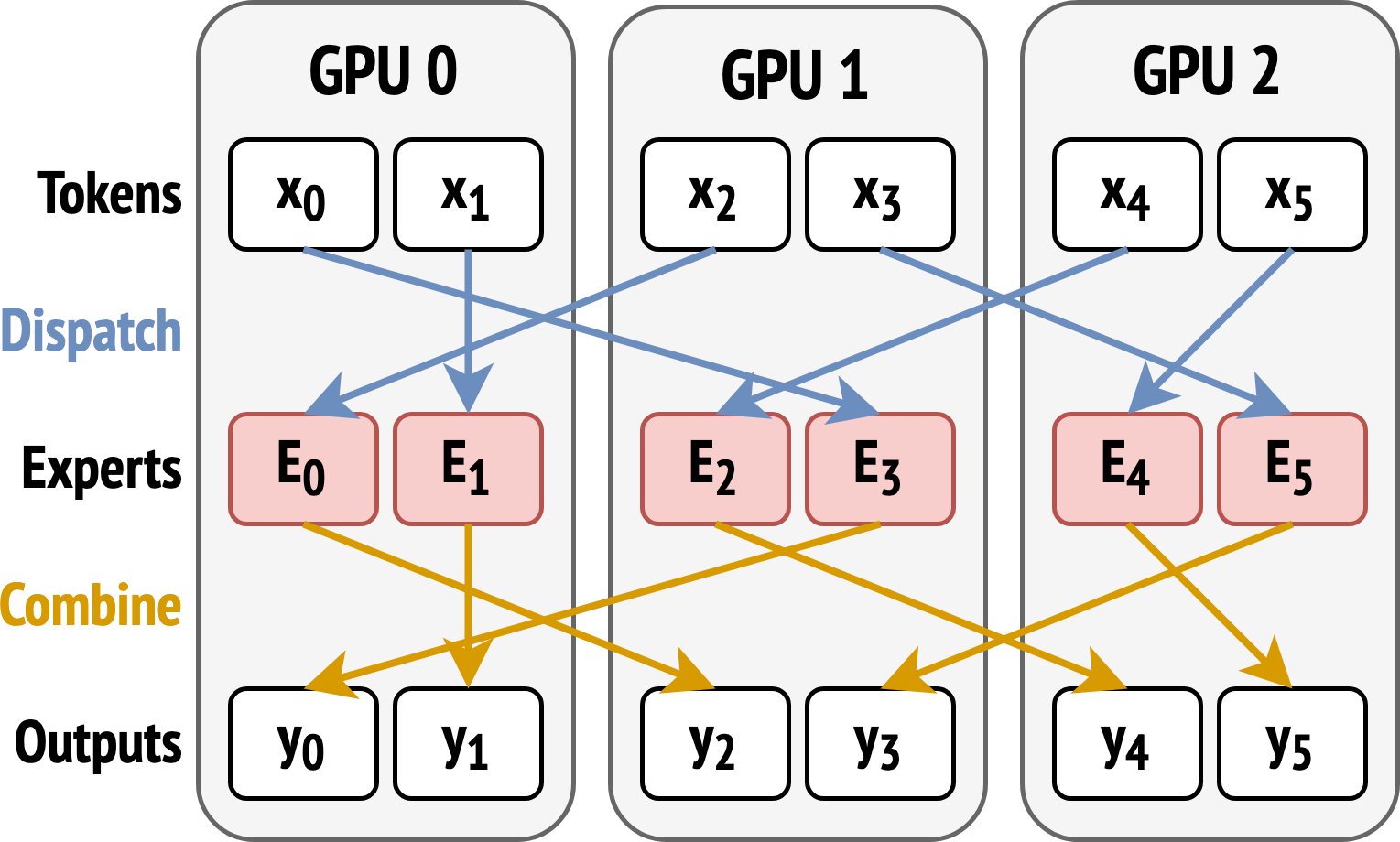}
    \vspace{-15pt}
    \caption{Expert parallelism example with an EP degree of 3 and top-1 routing.}
    \vspace{-10pt}
    \label{fig:expert-parallelism}
\end{wrapfigure}
For each input token, the MoE operation uses a router projection to select the top-\(k\) routed experts and assign each a router weight. The shared expert and each selected routed expert then apply a feed-forward network (FFN), consisting of up and gate projections, a gated activation, and a down projection. MoE sums the outputs, weighted by the normalized router weights:\footnote{Here, we consider the DeepSeek V3-style MoE architecture \cite{deepseekai2024deepseekv3,dai2024deepseekmoe}, but the principles generalize to other MoE architectures.}
\[
    \operatorname{MoE}(x, s) = f_{\mathrm{shared}}(x) + \sum_{i \in \mathcal{E}} \frac{s_i}{\sum_{j \in \mathcal{E}} s_j}\, f_i(x),
\]
where $x \in \mathbb{R}^{D}$ is the input token with hidden dimension $D$, $\mathcal{E}$ is the set of the $k$ selected routed experts, $s \in \mathbb{R}^{|\mathcal{E}|}$ are their router weights, and $f_i(x) = W_{\mathrm{down}}^{(i)} \bigl( \phi( W_{\mathrm{gate}}^{(i)} x ) \odot W_{\mathrm{up}}^{(i)} x \bigr)$ is the FFN of expert $i$, with activation function $\phi$, intermediate dimension $I$, and weights $W_{\mathrm{up}}^{(i)}, W_{\mathrm{gate}}^{(i)} \in \mathbb{R}^{I \times D}$ and $W_{\mathrm{down}}^{(i)} \in \mathbb{R}^{D \times I}$.

\textit{Expert parallelism} (EP) \cite{lepikhin2021gshard,fedus2022switch} partitions the routed experts across multiple GPUs, or \textit{ranks}. We call the number of ranks that collectively hold all expert weights the \textit{EP degree}. Figure~\ref{fig:expert-parallelism} shows an example: with 6 routed experts and an EP degree of 3, each rank holds 2 routed experts. Tokens must therefore be transferred across GPUs before and after the feed-forward computation, according to the router selection. We refer to the all-to-all transfer before the expert FFN as \textit{dispatch}, and the one after as \textit{combine}. 

Many systems optimize EP by \textit{overlapping} dispatch and combine with expert FFNs \cite{zhao2025deepep,zhang2025comet,aimuyo2025flashmoe,yu2026hybridep}. These schemes pipeline the token flow by transferring one chunk of tokens while computing the FFN on another. MoK is one variant of these schemes.

\subsection{GPUs and Scale-Up Architectures}
\label{sec:gpus-and-scale-up-architectures}

At a high level, GPUs load data from local or remote HBM, perform computation, and write the results back to local or remote HBM. To do this, a GPU kernel executes tens of thousands of hardware threads across more than a hundred \textit{streaming multiprocessors} (SMs), which can issue instructions to different execution units concurrently. Because these units saturate at different rates, achieving peak performance depends on overlapping their use to hide non-critical operations while maximizing the throughput of critical ones.

The GPU memory hierarchy trades latency for capacity as it moves farther from the SMs. Each SM contains 65,536 32-bit registers, accessible every clock cycle, and 228~KB of on-chip shared memory ($\approx$40~TB/s). All SMs share a 129~MB L2 cache ($\approx$22~TB/s) backed by 279~GB of HBM (8~TB/s). Threads can also access a remote GPU's HBM over NVLink (900~GB/s unidirectional) or InfiniBand/RoCE (100~GB/s unidirectional).\footnote{Here, we use the numbers for the Nvidia GB300 NVL72 \cite{nvidia2025gb300nvl72} as a representative example.}

\textit{Scale-up architectures} like NVL72 extend this hierarchy by connecting far more GPUs through a high-bandwidth, non-blocking fabric with uniform, single-hop any-to-any connectivity. This changes the tradeoff for parallelism strategies like EP. Parallelism axes that previously had to traverse the lower-bandwidth scale-out network can now stay entirely within the scale-up domain, raising the performance ceiling and enabling more aggressive overlapping.

\subsection{Related Work}
\label{sec:related-work}

\paragraph{Computation-Communication Overlap.} A large body of work overlaps GPU computation and communication for various AI workloads \cite{he2024asynctp,chang2024flux,liu2024ringattention,zhao2025deepep,zhang2025comet,aimuyo2025flashmoe,sul2026parallelkittens,zheng2025tritondist,thakkar2017cutlass,shoeybi2019megatron}. Some systems overlap host-triggered inter-GPU memory transfers with device kernels \cite{he2024asynctp,chang2024flux,zhao2025deepep,thakkar2017cutlass}, while others propose on-device schedulers that issue fine-grained, device-initiated communication \cite{zhang2025comet,aimuyo2025flashmoe}. Some works instead provide general abstractions and programming primitives that enable flexible overlap strategies \cite{zheng2025tritondist,sul2026parallelkittens}. These techniques have been applied to individual operators such as attention \cite{liu2024ringattention} and MoE \cite{zhang2025comet}, as well as to entire models \cite{shoeybi2019megatron,spector2025megakernel}. MoK builds on this body of work, specifically targeting distributed MoE training on scale-up platforms.

\paragraph{Efficient MoE Kernels.} MoK is inspired by three classes of systems that optimize MoE execution. First, some systems accelerate expert FFNs via block-sparse reformulation \cite{gale2023megablocks}, fused token permutation \cite{tan2024scattermoe}, low-precision expert-grouped GEMM \cite{zhao2025deepgemm}, or IO-efficient algorithmic redesign \cite{guo2026sonicmoe}. MoK reuses some of these techniques for local computation, such as SonicMoE's router gradient fusion (Section~\ref{sec:additional-features-and-optimizations}).

Second, many systems optimize EP communication and its overlap with computation via hierarchical all-to-all \cite{rajbhandari2022deepspeedmoe}, adaptive pipelining of all-to-all with expert FFNs \cite{hwang2023tutel}, dedicated dispatch and combine kernels designed for overlap \cite{zhao2025deepep}, or fine-grained device-initiated communication fused with expert FFNs \cite{zhang2025comet,aimuyo2025flashmoe}. Closest to MoK is MegaMoE, a fused EP megakernel in the DeepGEMM library \cite{zhao2025deepgemm}. However, MegaMoE is an inference-only forward kernel specialized for DeepSeek-V4 serving in FP8$\times$FP4, whereas MoK supports full training on scale-up architectures with training-preferred precisions like BF16 and MXFP8.

Third, several end-to-end frameworks \cite{shoeybi2019megatron,rasley2020deepspeed,nvidia2023tensorrtllm,kwon2023vllm,zheng2024sglang,zhu2025nanoflow} integrate efficient per-operator kernels and optimization techniques into full training or inference stacks. MoK is complementary to these systems and can serve as their MoE backend, as it already does for Nvidia NeMo AutoModel \cite{nvidia2025automodel}.

\paragraph{Megakernels.} MoK adopts the megakernel approach \cite{spector2025megakernel,spector2025wholegpu,cheng2026mpk}, in which a single persistent kernel executes multiple operators to remove per-kernel launch latency, achieve inter-operator overlap, and minimize straggler tasks. In MoK, dispatch, expert FFNs, and combine are integrated into a single megakernel. Megakernels, however, only provide a mechanism for combining kernels at operator boundaries and do not prescribe operator fusion or computation-communication overlap, which is where MoK's main design decisions lie.

\section{Mixture-of-Kittens}
\label{sec:mixture-of-kittens}

\begin{wrapfigure}{r}{0.5\textwidth}
    \centering
    \vspace{-40pt}
    \includegraphics[width=0.48\textwidth]{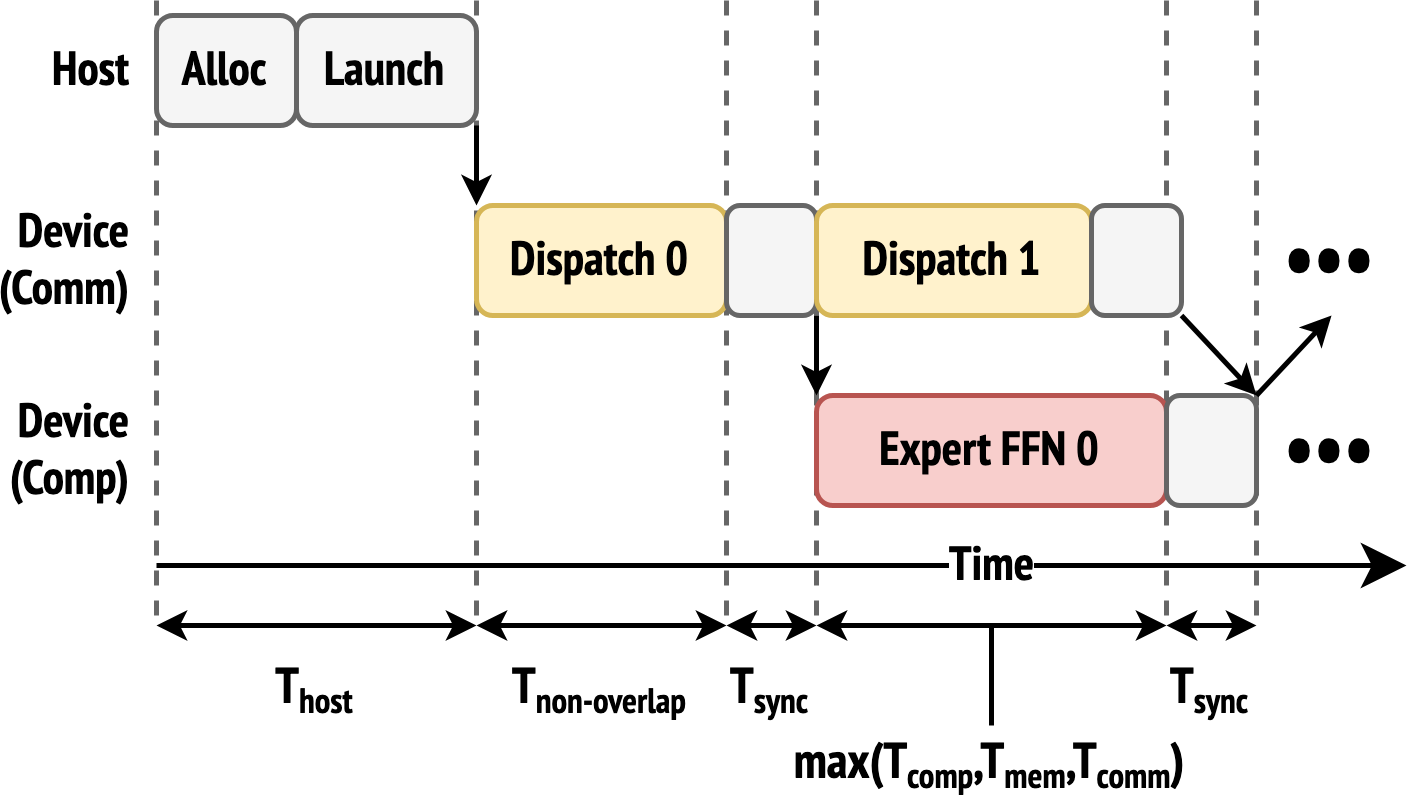}
    \vspace{-10pt}
    \caption{Cost model terms on a naive overlapped MoE execution. Unlabeled gray boxes denote synchronization. Arrows mark execution dependencies.}
    \label{fig:cost-model}
    \vspace{-20pt}
\end{wrapfigure}
MoK's goal is to minimize end-to-end MoE latency \(C\). At a high level, this decomposes as:
\begin{equation}
    \begin{split}
        C = {} & \max(T_{\text{comp}},\, T_{\text{mem}},\, T_{\text{comm}}) \\
               & + T_{\text{non-overlap}} + T_{\text{sync}} + T_{\text{host}}.
    \end{split}
    \label{eq:cost-model}
\end{equation}

Here, \(T_{\text{comp}}\), \(T_{\text{mem}}\), and \(T_{\text{comm}}\) denote computation, intra-GPU memory access, and inter-GPU communication times, respectively. If these components overlap perfectly, their joint cost is their maximum. \(T_{\text{non-overlap}}\) accounts for the non-overlapped portions of these operations, such as the initial dispatch and the final combine. \(T_{\text{sync}}\) covers the intra- and inter-GPU synchronization time spent on completion signals and memory fences. \(T_{\text{host}}\) captures host-side overheads like CPU-GPU synchronization, memory allocation, and kernel launch latency. Figure~\ref{fig:cost-model} illustrates these terms on a simple overlapped MoE execution.

Each of \(T_{\text{comp}}\), \(T_{\text{mem}}\), and \(T_{\text{comm}}\) must also be minimized individually by driving its hardware unit to peak utilization. For \(T_{\text{comp}}\), this means keeping the tensor cores continuously fed by pipelining data operands. For \(T_{\text{comm}}\), it means saturating the network lanes with outstanding transfers and spreading traffic evenly.

The remainder of this section presents three design choices that minimize \(C\). In addition, Section~\ref{sec:additional-features-and-optimizations} discusses other features and optimizations that further improve production readiness and performance.

\subsection{Choosing the Right Communication Direction}
\label{sec:choosing-the-right-communication-direction}

\begin{figure}[t]
    \vspace{-5pt}
    \centering
    \begin{minipage}[c]{0.46\textwidth}
        \centering
        \includegraphics[width=\linewidth]{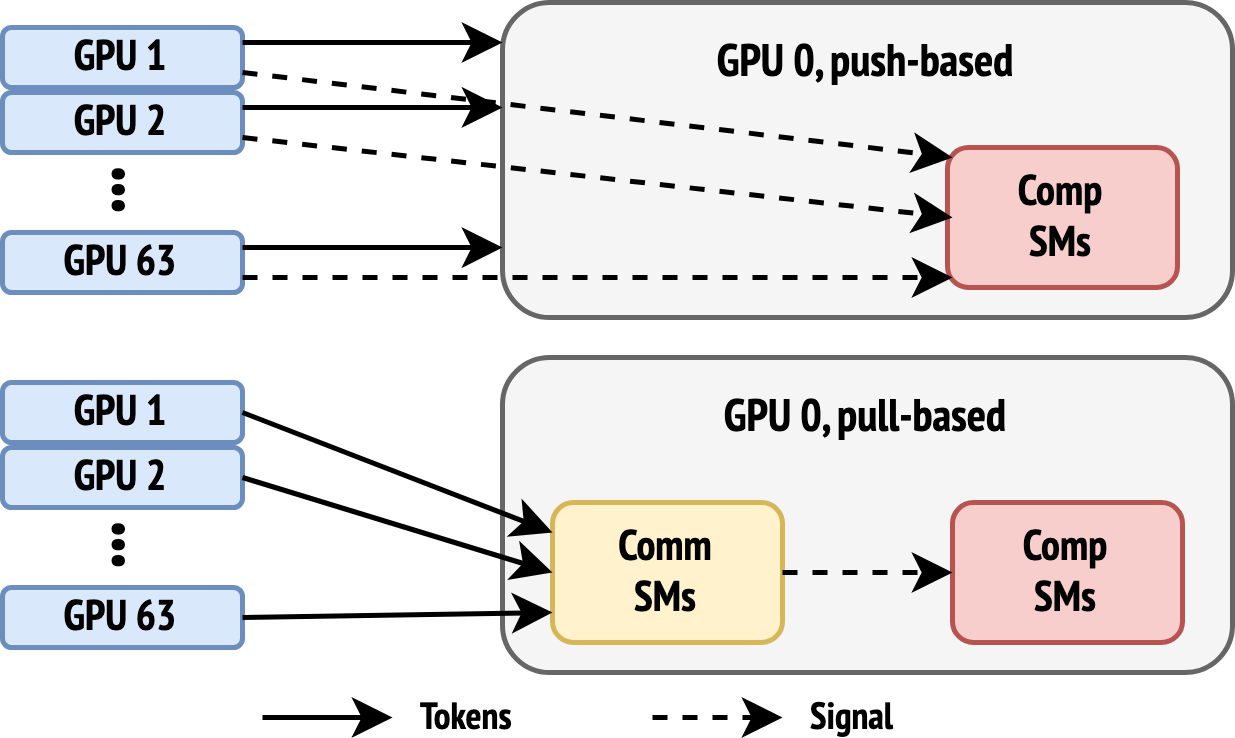}
    \end{minipage}\hfill
    \begin{minipage}[c]{0.50\textwidth}
        \centering
        \small
        \setlength{\tabcolsep}{5pt}
        \renewcommand{\arraystretch}{1.15}
        \begin{tabular}{@{}rcccc@{}}
            \toprule
            & \multicolumn{2}{c}{Full latency ($\mu$s)} & \multicolumn{2}{c}{Signal overhead ($\mu$s)} \\
            \cmidrule(lr){2-3}\cmidrule(l){4-5}
            EP degree & Pull & Push & Pull & Push \\
            \midrule
            4  & 475 & 489 &  1 &  39 \\
            8  & 559 & 574 &  1 &  51 \\
            16 & 603 & 633 &  4 &  66 \\
            32 & 638 & 704 & -1 & 119 \\
            64 & 638 & 744 & -2 & 135 \\
            \bottomrule
        \end{tabular}
    \end{minipage}
    \vspace{-5pt}
    \caption{Signaling under push and pull dispatch. Left: with push, GPU 0's computation SMs wait for a completion signal from every peer; with pull, they only need a GPU-local signal from their own communication SMs. Right: dispatch latency and signaling overhead on NVL72 (7,168 BF16 tokens of width 7,168 per GPU, 4 local experts, 4 uniformly random routes per token). Signal overhead is the added latency of in-kernel completion signaling relative to the same kernel without it.}
    \vspace{-5pt}
    \label{fig:push-pull-dispatch}
\end{figure}

This section explains why communication direction matters for fine-grained MoE overlap, which direction MoK chooses for each operator, how we optimize the chosen direction, and what further benefits it brings.

Under fine-grained computation-communication overlap, signaling (\(T_{\text{sync}}\)) becomes a bottleneck, because each task (e.g., a tensor-core matrix multiply) cannot begin until its input data is known to have arrived. The direction of communication determines how much signaling is required. Figure~\ref{fig:push-pull-dispatch} (left) illustrates this for dispatch. With push-based dispatch, every peer must send a completion signal each time it pushes a chunk of tokens, whereas with pull-based dispatch a local signal suffices once the destination GPU has issued all of its loads. As Figure~\ref{fig:push-pull-dispatch} (right) shows, push-based dispatch spends 39--135\,\(\mu\)s on signaling, or 8--18\% of dispatch time, and this cost grows with EP degree. On the other hand, pull-based signaling stays within 4\,\(\mu\)s. A fine-grained megakernel incurs this overhead many times per execution, so the cost compounds.

Prior EP MoE systems predominantly use push-based transfers \cite{zhao2025deepep,mao2026uep,yu2026hybridep}, since push outperforms pull for fine-grained RDMA transfers \cite{wei2018hybridrdma,ma2026demystifyingnvshmem}. On scale-up platforms, however, the entire EP group can fit within a scale-up domain, which frees us to choose each MoE operator's communication direction independently.

MoK therefore chooses the direction of each operator so that the GPU receiving the signal also issues the remote loads or stores. For the forward pass, each GPU pulls its inputs (pull-based dispatch), signals its local computation workers, and pushes its outputs (push-based combine) once they signal completion. The backward pass mirrors this with pull-based reverse-combine and push-based reverse-dispatch. As a result, the handoff between communication and computation never leaves the GPU, eliminating cross-GPU signaling.

\begin{algorithm}[t]
    \caption{Pull-based dispatch schedule construction}
    \label{alg:dispatch-schedule-construction}
    \begin{algorithmic}
        \Require $R$ ranks, each with $N$ tokens and $L$ experts; top-$K$ routes $A \in \{0, \dots, RL-1\}^{R \times NK}$, where $A_{r,j}$ is the expert chosen by route $j$ on source rank $r$; current rank $c$; local expert $\ell$ is global expert $cL+\ell$
        \Ensure schedule $\Phi$, where $\Phi_p = (r, j)$ means local buffer position $p$ receives route $j$ from source rank $r$
        \State $M_{r, \ell} \gets \bigl|\{\, j : A_{r,j} = cL + \ell \,\}\bigr| \quad \forall\, r, \ell$ \Comment{route count per (source rank, local expert)}
        \State $S_{\ell} \gets \sum_{\ell' < \ell} \sum_{r} M_{r,\ell'} \quad \forall\, \ell$ \Comment{start of expert $\ell$'s buffer region}
        \ForAll{local expert $\ell$ and routes $(r, j)$ with $A_{r,j} = cL + \ell$}
            \State $o \gets \bigl|\{\, j' < j : A_{r,j'} = A_{r,j} \,\}\bigr|$
            \State $p \gets \sum_{r'=0}^{R-1} \min(M_{r',\ell},\, o) + \bigl|\{\, r' < r : M_{r',\ell} > o \,\}\bigr|$
            \State $\Phi_{S_{\ell} + p} \gets (r,\ j)$
        \EndFor
    \end{algorithmic}
\end{algorithm}

\paragraph{Scheduling.} Once the communication direction is chosen, we must decide the order of token transfers, so that it both saturates the network links (\(T_{\text{comm}}\)) and keeps the receiving tensor cores busy (\(T_{\text{comp}}\)).

Let us consider pull-based dispatch first. At each step, a GPU must decide (1) which peer to pull from, (2) which token to pull from that peer, and (3) where in local memory to store it. For (1), peers should be selected round-robin so that all network lanes carry roughly the same number of outstanding requests. For (2), tokens should arrive in non-decreasing local-expert order so that the per-expert matrix multiplies do not become fragmented. For (3), tokens should fill the local buffer contiguously so that no additional local memory movement is needed before they are fed to the tensor cores.

MoK satisfies all three conditions through a separate kernel that builds a schedule table, which is then passed to the megakernel (Figure~\ref{fig:pull-figure}, right). The table has two columns, \texttt{\{src\_rank, src\_index\}}, and its row index is the token's position in the destination buffer. Dispatch simply walks the table row by row. Algorithm~\ref{alg:dispatch-schedule-construction} details how the table is built. Conceptually, it sorts the routes by the key (local expert $\ell$, per-source-rank ordinal $o$, source rank $r$), where $o$ counts the earlier routes from the same source rank to the same expert. Sorting by $\ell$ first gives each expert a contiguous region, and sorting by $(o, r)$ within a region interleaves the source ranks round-robin. MoK's GPU implementation parallelizes this and builds the schedule in 2.1--7.7\% of MoE runtime across our workloads, a 1.4--2.2$\times$ smaller overhead than Comet's~\cite{zhang2025comet} (Appendix~\ref{app:scheduling-overhead}). We include this time in our benchmarks (Section~\ref{sec:experiments}).

Finally, the schedule can be built once and reused by all four communication operators. Push-based combine, for instance, can reinterpret \texttt{\{src\_rank, src\_index\}} as \texttt{\{dst\_rank, dst\_index\}}, and the pushes will remain evenly spread across peers. The table occupies only a few megabytes in the worst case, so retaining it across operators costs negligible memory.


\subsection{Restructuring the Computation-Communication Overlap}
\label{sec:restructuring-computation-communication-overlap}

This section describes computation-communication overlap granularity, shows that the optimal granularity depends on the workload, and explains how MoK supports the full range of granularities efficiently.

Overlapping computation and communication requires exchanging many readiness signals, one for each chunk of tokens transferred and computed. We call this chunk a \textit{minibatch}, and its size sets the \textit{granularity} of the overlap. At one extreme, the overlap can be very fine-grained, at just enough tokens to fill a single tensor-core matrix-multiply-accumulate instruction, as in Comet~\cite{zhang2025comet}. At the other, it can be coarse-grained, at tens of thousands of tokens. This is the common usage of DeepEP~\cite{zhao2025deepep}.

Neither extreme is optimal. Let \(N_{\mathrm{routed}}\) denote the number of routed tokens, \(b\) the minibatch size, and \(w\) the workload shape, including the hidden and intermediate dimensions. A small \(b\) never fully saturates the tensor cores and pays a per-minibatch cost \(C_{\mathrm{ineff}}(b,w)\) of underutilized compute and synchronization overhead, giving a fine-grained penalty \(P_{\mathrm{fine}}(b,w) \approx \lceil N_{\mathrm{routed}}/b \rceil\, C_{\mathrm{ineff}}(b,w)\). A large \(b\) leaves the dispatch of the first minibatch and the combine of the last minibatch unoverlapped, giving a coarse-grained penalty \(P_{\mathrm{coarse}}(b) \approx T_{\mathrm{dispatch}}(b) + T_{\mathrm{combine}}(b)\). Increasing \(b\) lowers the first term and raises the second, so \(b\) should be chosen per workload to minimize their sum, \(P_{\mathrm{fine}}(b,w) + P_{\mathrm{coarse}}(b)\).

\begin{figure}[t]
    \vspace{-25pt}
    \centering
    \begin{minipage}[c]{0.58\textwidth}
        \centering
        \includegraphics[width=\linewidth]{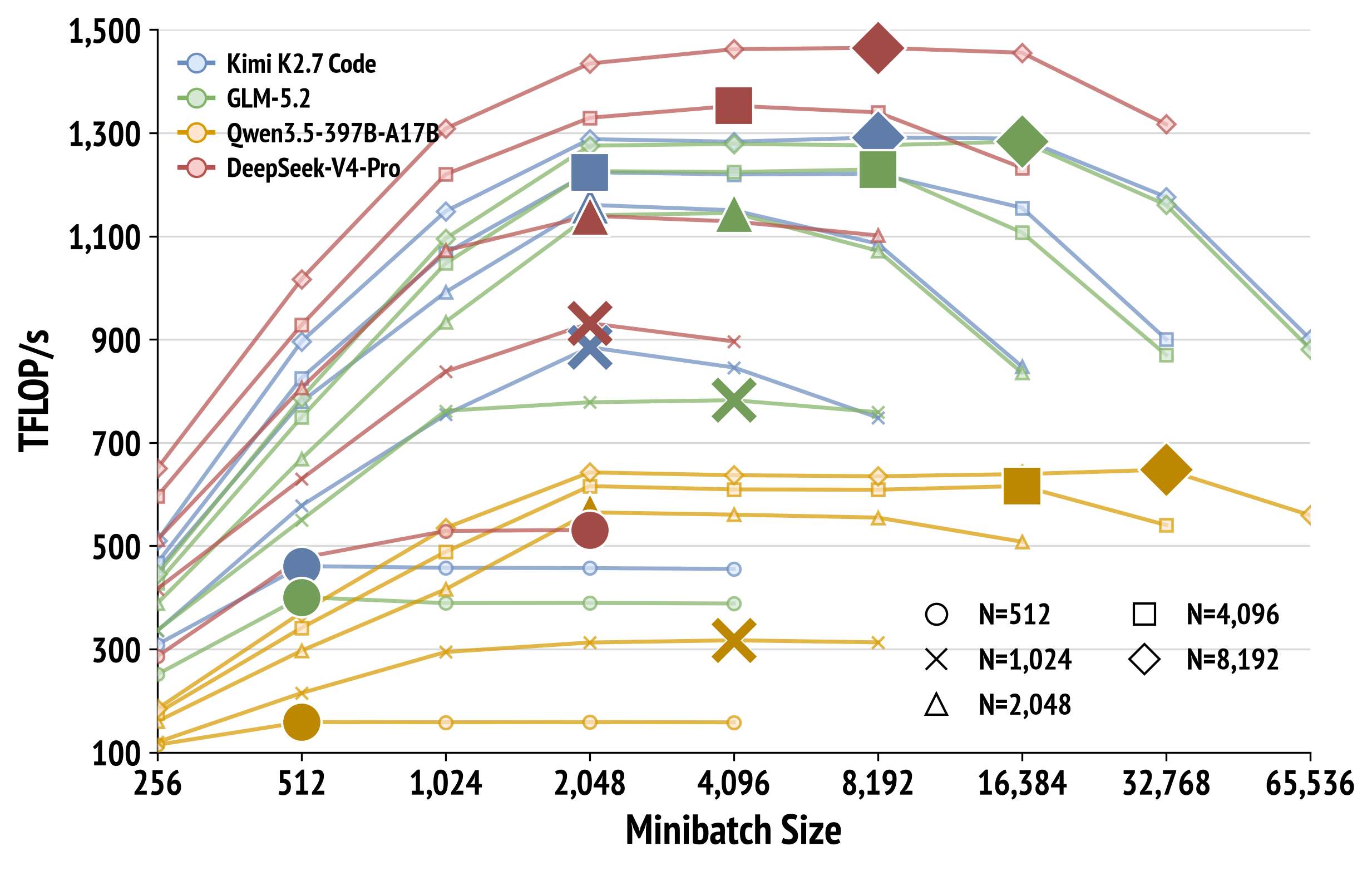}
    \end{minipage}\hfill
    \begin{minipage}[c]{0.38\textwidth}
        \centering
        \includegraphics[width=\linewidth]{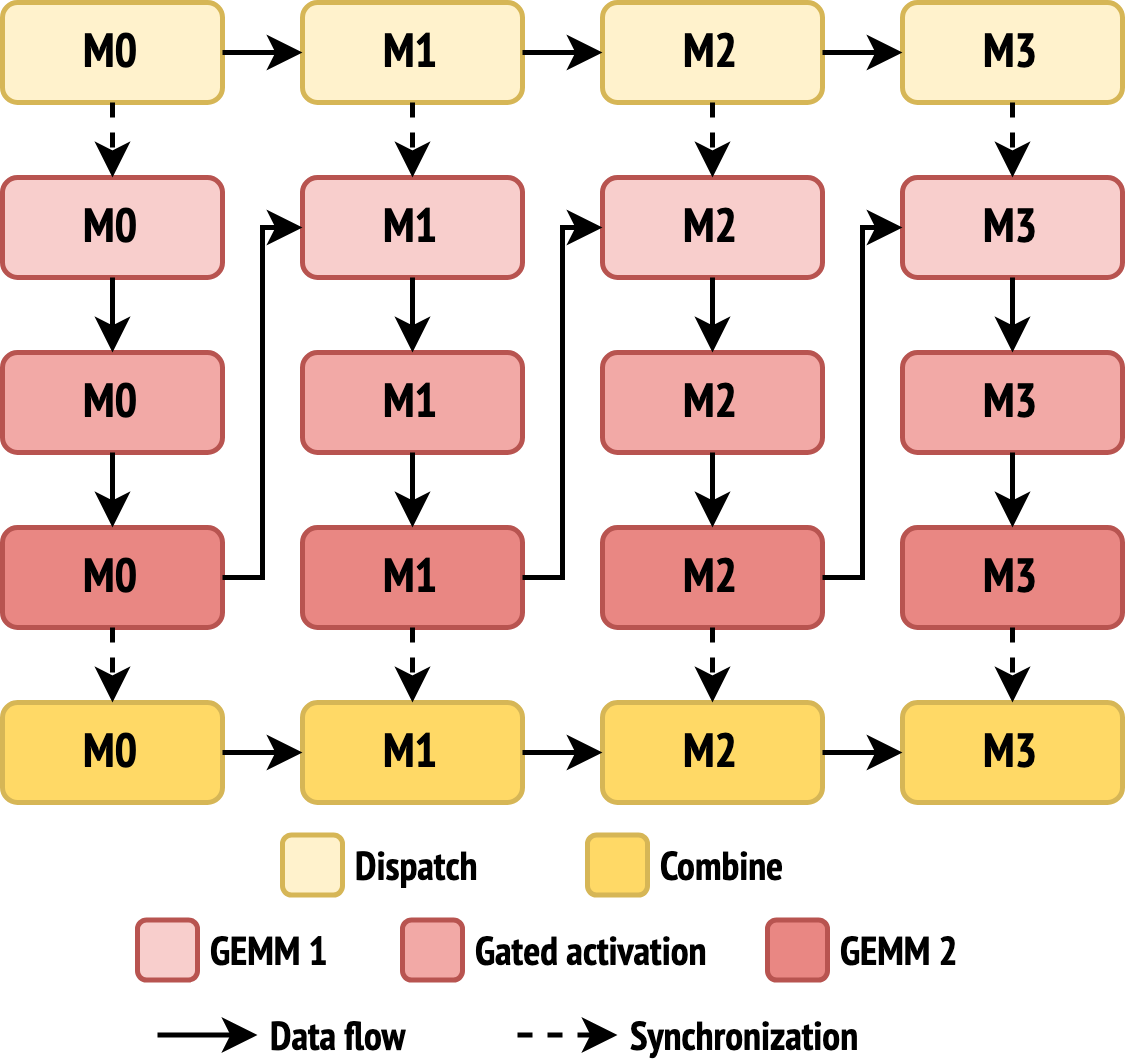}
    \end{minipage}
    \vspace{-5pt}
    \caption{Left: MoK MXFP8 forward-pass throughput on 16 GPUs as a function of minibatch size. Colors denote model shapes, symbols denote the number \(N\) of input tokens per GPU, and enlarged symbols mark the throughput-maximizing minibatch size for each curve. For a fixed model and token count, the best minibatch size achieves \(1.38\times\)--\(3.53\times\) the throughput of the worst. Right: maximally deferred synchronization with four minibatches (M0--M3). Communication (yellow) and computation (red) traverse the work in different orders and meet only at the dashed arrows.}
    \vspace{-5pt}
    \label{fig:overlap-granularity}
\end{figure}

Figure~\ref{fig:overlap-granularity} (left) shows this tradeoff empirically. Across the evaluated workloads, throughput first rises with minibatch size as tensor-core utilization improves and per-minibatch overhead amortizes. It then plateaus or declines as coarser minibatches expose communication and tensor-core bubbles. The throughput-maximizing minibatch size ranges from 512 to 32,768 tokens across model shapes and token counts.

Thus, an MoE system should not fix one granularity at design time. Instead, it should make granularity a first-class tunable parameter and keep its overhead low at any granularity. MoK achieves this by (1) maximally deferring the computation-communication synchronization and (2) fusing all computation and communication into a single megakernel.

\paragraph{Maximally Deferred Synchronization.} MoK lets computation and communication run as far ahead as the data dependencies allow. Figure~\ref{fig:overlap-granularity} (right) illustrates this. Communication proceeds \textit{breadth-first} over minibatches, dispatching all tokens continuously without waiting for computation to complete, whereas computation proceeds \textit{depth-first}, running one minibatch through the entire expert FFN before moving on to the next. The two sides signal each other asynchronously only at the dashed arrows, where computation waits for a minibatch to arrive and combine waits for its outputs.

\paragraph{Kernel Fusion.} MoK fuses all computation and communication into a single megakernel~\cite{spector2025megakernel} and runs them concurrently through inter-SM overlapping~\cite{sul2026parallelkittens}. Some SMs are assigned exclusively to the expert FFNs and others to dispatch and combine. The two groups signal each other through a GPU-local barrier. Without a megakernel, the schedule in Figure~\ref{fig:overlap-granularity} (right) would require launching several kernels per minibatch and overlapping them through multiple streams. This would incur kernel launch and stream synchronization costs (\(T_{\text{host}}\)) of tens of microseconds per minibatch, and it would leave the interleaving to the hardware scheduler, which does not guarantee concurrent execution of computation and communication~\cite{nvidia2026greencontexts}.

\subsection{Eliminating CPU-GPU Synchronization}
\label{sec:eliminating-cpu-gpu-synchronization}

This section quantifies the cost of CPU-GPU interaction on the NVL72 platform, describes MoK's token ring buffering, and explains how it stays efficient in both the forward and backward passes.

\begin{figure}[t]
    \vspace{-15pt}
    \centering
    \begin{minipage}[c]{0.55\textwidth}
        \centering
        \includegraphics[width=\linewidth]{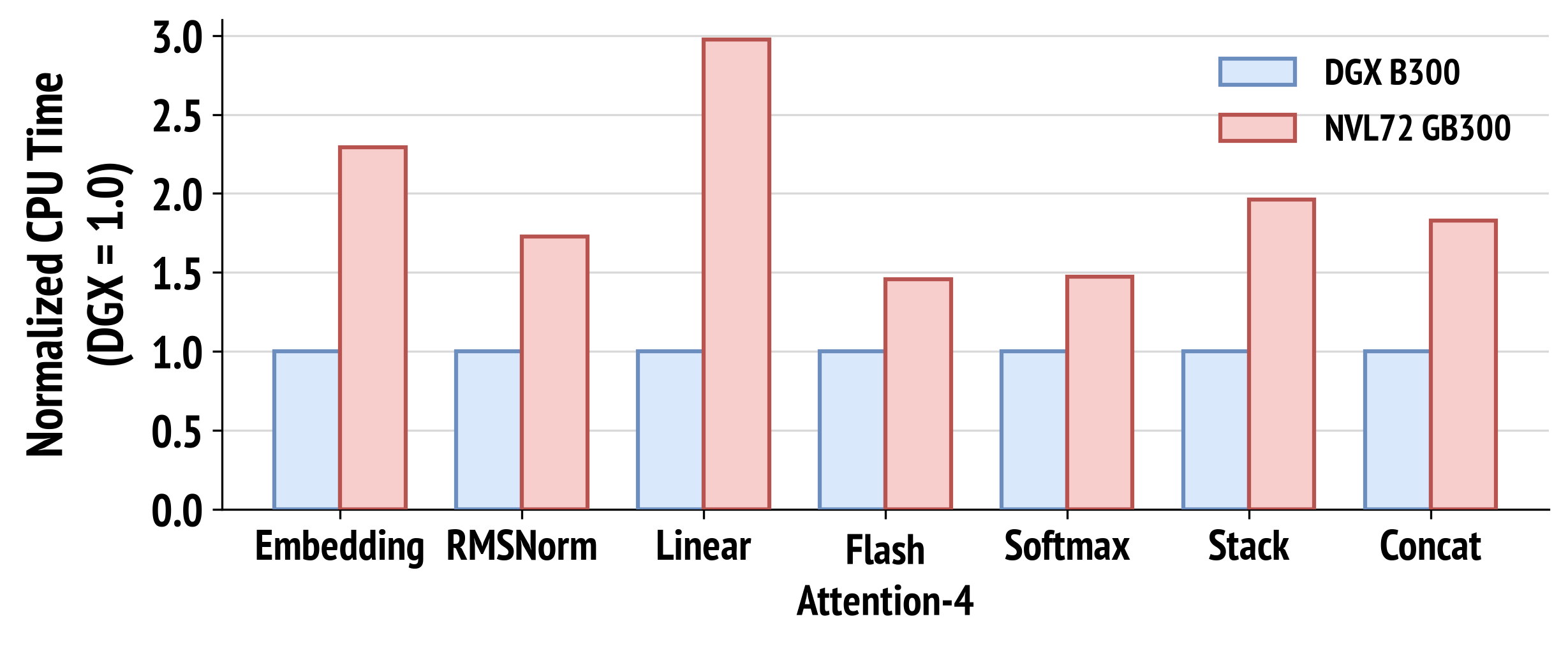}
    \end{minipage}\hfill
    \begin{minipage}[c]{0.40\textwidth}
        \centering
        \small
        \setlength{\tabcolsep}{5pt}
        \renewcommand{\arraystretch}{1.15}
        \begin{tabular}{@{}lrrr@{}}
            \toprule
            & \multicolumn{3}{c}{Local tokens per rank} \\
            \cmidrule(l){2-4}
            MoE Shape & 1{,}024 & 2{,}048 & 4{,}096 \\
            \midrule
            Kimi K2.7 Code    & 0.3\% & 1.2\% & 0.7\% \\
            GLM-5.2           & 7.4\% & 4.6\% & 1.9\% \\
            Qwen3.5-397B-A17B & 0.7\% & 1.3\% & 2.5\% \\
            DeepSeek-V4-Pro   & 0.1\% & $-$0.5\% & 1.0\% \\
            \bottomrule
        \end{tabular}
    \end{minipage}
    \vspace{-5pt}
    \caption{Left: host-side invocation time of common PyTorch operations on a GB300 NVL72, normalized to a DGX B300. The x-axis is the operation and the y-axis is the normalized CPU time. Right: latency overhead of MoK's macrobatch ring buffer. We compare a ring buffer that can hold half of the routed tokens against an over-allocated buffer that can hold all of them, and report the percentage increase in latency.}
    \vspace{-5pt}
    \label{fig:slow-cpu}
\end{figure}

Host-side execution on GB300 NVL72 systems is slow relative to conventional DGX systems~\cite{marks2025tecpuoverhead}. In Figure~\ref{fig:slow-cpu} (left), we measure the host-side invocation time of common PyTorch operations on both platforms, and find that the NVL72 host path is \(1.46\times\) to \(2.97\times\) slower. At production scale, we observe that GPU execution is thus far more likely to be bottlenecked by host work, and CPU-GPU synchronization becomes much more costly, since it prevents the GPU stream from running ahead of the CPU thread.

One common source of such synchronization in EP execution is memory allocation. Because the number of tokens routed to each expert is not known until runtime, allocating exact-sized destination buffers requires copying these counts to the CPU before the allocation~\cite{gale2023megablocks}. An alternative is to drop tokens that exceed a fixed buffer capacity~\cite{lepikhin2021gshard,fedus2022switch}, but this degrades training quality~\cite{gale2023megablocks}.

MoK avoids both CPU-GPU synchronization and token dropping by managing a token ring buffer entirely on the device. We call this ring buffer a \textit{macrobatch} and expose its size as a tunable parameter. The ring buffer enables a producer-consumer pipeline between dispatch and the expert FFNs, and between the expert FFNs and combine. Each producer writes into one region of the macrobatch while its consumer works on another. For example, in Figure~\ref{fig:pull-figure} (top left), dispatch writes tokens to the third slot while the expert FFNs process the second slot and combine reads from the first slot. The same pipelining continues when the ring wraps around, so the ring boundary adds almost no overhead. Compared to over-allocating a buffer that holds every routed token, which is impractical in production yet shows the best achievable performance, ring buffering adds only 1.7\% latency in geometric mean across our workloads (Figure~\ref{fig:slow-cpu}, right).

\paragraph{Ring-Aware Forward Replay.} In production training, device memory cannot hold all intermediate activations of the forward pass, so the backward pass must recompute some of them through forward replay. However, naively rerunning the MoK forward pass preserves only the activations of the last macrobatch, since ring buffering overwrites everything before it, and the forward megakernel recomputes activations that the backward pass never consumes (e.g., the down-projection output).

MoK therefore fuses a ring-aware forward replay directly into the backward megakernel, where it recovers only the missing intermediates and is overlapped with the backward pass. We further minimize this replay by maximizing the number of tokens that survive in the ring after the forward pass. If the forward pass processes macrobatches in ascending order, only the last macrobatch survives, which is usually partial unless all tokens fit in one macrobatch. MoK's forward pass instead processes tokens in \textit{reverse macrobatch order}, so that it ends with a completely full ring and the backward pass replays as little as possible.

\subsection{Additional Features and Optimizations}
\label{sec:additional-features-and-optimizations}

Beyond the three design choices described in this section, MoK implements additional features and optimizations that are essential for production MoE training.

\paragraph{Determinism.} To enable training ablations and on-policy reinforcement learning, MoK is fully deterministic. The order of floating-point operations is fixed across executions, so the same input produces bitwise-identical output regardless of hardware scheduling. The main source of nondeterminism in GPUs is atomic operations. MoK either forces atomics to the same address to execute in a fixed order through software scheduling, or replaces them with writes to a temporary buffer that another worker reduces in a deterministic order.

\paragraph{RDMA Overlap via Cluster Launch Control.} MoK schedules its tasks through Cluster Launch Control (CLC) \cite{nvidia2026clusterlaunchcontrol}, a hardware-native work-stealing mechanism introduced in Nvidia Blackwell. A CLC-based kernel can naturally yield SMs to kernels on a higher-priority stream without having to fully complete. This matters because, during training, we often need to overlap intra-rack communication and computation with inter-rack communication, such as FSDP all-gathers, which needs SMs of its own. Without CLC, that inter-rack communication would serialize behind the megakernel, causing performance degradation.

\paragraph{MXFP8 Support.} For better hardware utilization, MoK supports both BF16 and MXFP8 \cite{rouhani2023microscaling} execution. In MXFP8 mode, the shared expert FFN stays in BF16 for training stability. One added cost of MXFP8 is that tensors must be quantized before they reach the tensor cores. To minimize this overhead, MoK (1) provides an optimized MXFP8 quantization kernel for pre-quantizing the weights, and (2) fuses activation quantization into the dispatch, the expert-grouped GEMMs, and the gated activation computation.

\paragraph{Router Weight Gradient Computation.} The MoK backward also computes the router weight gradients, using the SonicMoE-style calculation~\cite{guo2026sonicmoe}. Rather than materializing the down-projection output, MoK computes the router weight gradients from the inner product of the gated activation and the down-projection input gradient. This computation is fused into the gated activation backward to avoid additional HBM traffic.

\paragraph{Shared Expert.} Many open-weight MoE architectures include a shared expert that every token passes through regardless of routing. The shared expert's load is uniform across EP ranks and it does not depend on dispatch or combine. Therefore, MoK fits the shared expert FFN in the first dispatch window, where the computation SMs would otherwise sit idle.

\paragraph{Computation-Communication SM Tuning.} The optimal split between computation and communication SMs depends on the workload~\cite{sul2026parallelkittens}. Local token count, model shape, and hardware compute and network bandwidth all change how long dispatch and combine take relative to the expert FFN tasks. Even within the same workload, the optimum differs between the forward and backward passes. MoK therefore exposes the number of communication SMs as a tunable parameter separately for the forward and backward passes.

\section{Experiments}
\label{sec:experiments}

Section~\ref{sec:mixture-of-kittens} introduced MoK and its three key design choices: communication direction, overlap granularity, and ring buffering. To evaluate their effectiveness, we benchmark MoK at two levels: (1) the effective compute throughput of standalone MoE forward and backward passes (Section~\ref{sec:moe-layer-benchmarks}), and (2) the end-to-end training throughput on a production training stack with 512 GPUs (Section~\ref{sec:End-to-end-benchmarks}). All experiments were run on GB300 NVL72 racks, each containing 72 B300 GPUs fully interconnected via fifth-generation NVLink~\cite{nvidia2025gb300nvl72}, using CUDA 13.0, Python 3.13, and PyTorch 2.13. We made a best-effort attempt to tune every baseline for maximum performance, and all final configurations can be inspected in our open-source repository.

\subsection{MoE Layer Benchmarks}
\label{sec:moe-layer-benchmarks}

We benchmark the complete execution of a single MoE layer, including global token scheduling, dispatch, routed and shared expert FFNs, combine, and the router-weighted sum. Router logits and model inputs are generated from a standard normal distribution, and every implementation receives bitwise-identical copies of them. We measure the forward and backward passes separately in both BF16 and MXFP8, timing 100 executions after 500 warmup iterations.

We take the latency of the slowest rank for each execution, record the median over the 100 executions, and convert it to effective expert-FFN throughput. For $N$ input tokens, hidden dimension $D$, expert intermediate dimension $I$, top-$k$ routing, and measured time $t$, forward throughput is $6N(k+1)DI/t$, where $k+1$ accounts for the $k$ routed experts and one shared expert. The backward pass uses twice this FLOP count.

\begin{figure}[t]
    \vspace{-20pt}
    \centering
    \includegraphics[
        width=1.0\textwidth
    ]{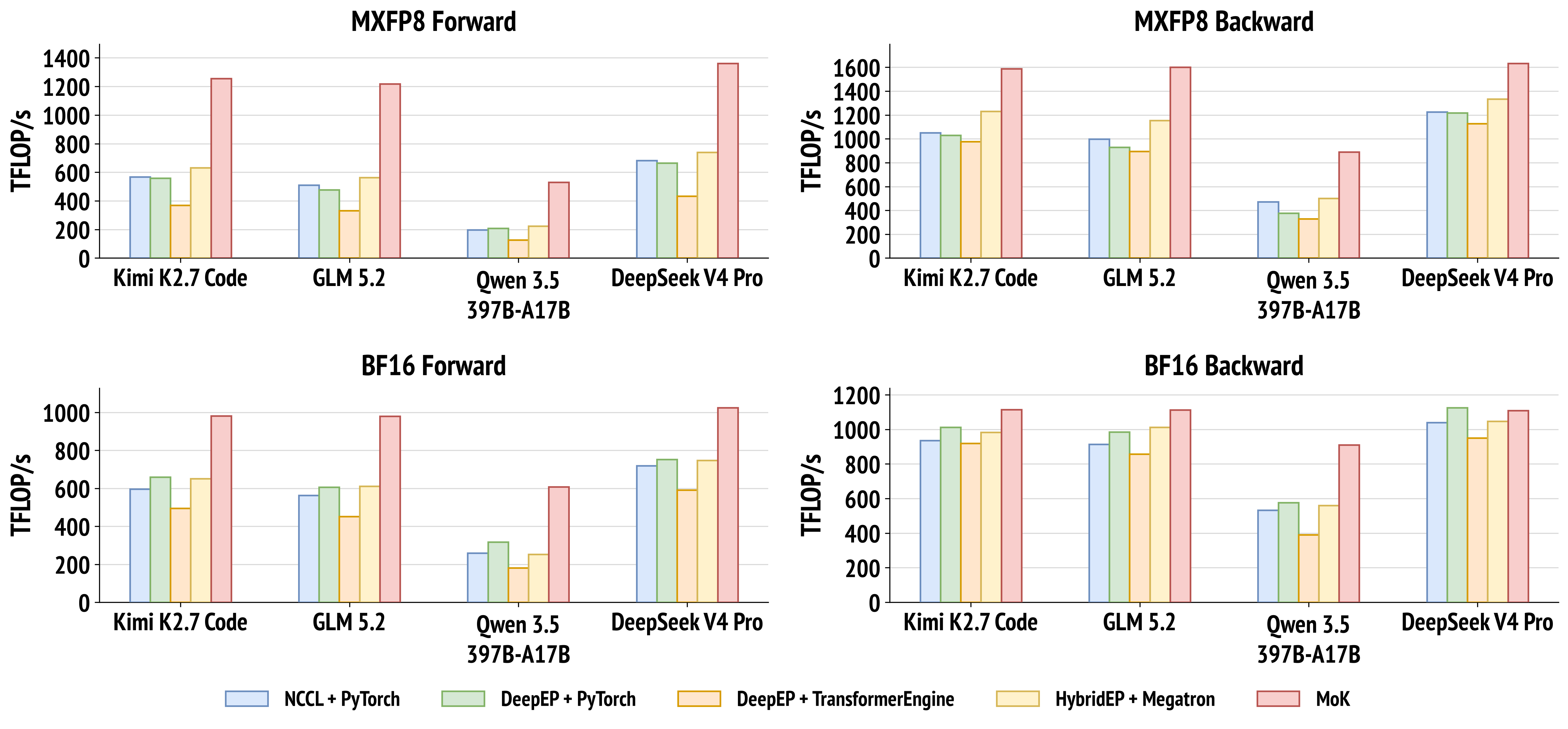}
    \vspace{-20pt}
    \caption{MoE layer throughput per GPU on a GB300 NVL72. The x-axis is the MoE shape and the y-axis is the effective expert-FFN throughput in TFLOP/s. We use an EP degree of 64 and 2{,}048 input tokens per GPU before routing. Higher is better.}
    \vspace{-5pt}
    \label{fig:main-benchmarks}
\end{figure}

We compare against the publicly available baselines that (1) support both MoE forward and backward passes and (2) support execution on NVL72 systems:

\begin{itemize}[itemsep=0pt, topsep=0pt]
    \item NCCL~\cite{nvidia2015nccl} + PyTorch~\cite{paszke2019pytorch}
    \item DeepEP~\cite{zhao2025deepep} + PyTorch
    \item DeepEP + TransformerEngine~\cite{nvidia2022transformerengine}
    \item HybridEP~\cite{yu2026hybridep}\footnote{Megatron also supports DeepEP, but Nvidia recommends HybridEP for optimal performance on NVL72.} + Megatron~\cite{shoeybi2019megatron}
\end{itemize}

We evaluate the MoE layer shapes of four widely used open-weight models, each with one shared expert, $E$ routed experts, hidden dimension $D$, expert intermediate dimension $I$, and $k$ routes per token:

\begin{itemize}[itemsep=0pt, topsep=0pt]
    \item Kimi K2.7 Code~\cite{kimi2026k27code} ($E{=}384$, $D{=}7168$, $I{=}2048$, top-$k{=}8$)
    \item GLM-5.2~\cite{glm5team2026glm5} ($E{=}256$, $D{=}6144$, $I{=}2048$, top-$k{=}8$)
    \item Qwen3.5-397B-A17B~\cite{qwen2026qwen35} ($E{=}512$, $D{=}4096$, $I{=}1024$, top-$k{=}10$)
    \item DeepSeek-V4-Pro~\cite{deepseekai2026v4} ($E{=}384$, $D{=}7168$, $I{=}3072$, top-$k{=}6$)
\end{itemize}

Figure~\ref{fig:main-benchmarks} shows MoE layer throughput for every combination of model shape, pass (forward and backward), and precision (MXFP8 and BF16), using an EP degree of 64 and 2{,}048 tokens per GPU before dispatch. Relative to the fastest baseline for each combination, MoK achieves speedups of up to 2.37$\times$ for MXFP8 forward, 1.78$\times$ for MXFP8 backward, 1.92$\times$ for BF16 forward, and 1.58$\times$ for BF16 backward.

We observe that MoK's relative gains are larger for MXFP8 than for BF16, and larger for the forward pass than for the backward pass. This is because peak BF16 tensor-core throughput on GB300 is half that of MXFP8 \cite{nvidia2025gb300nvl72}, and the backward pass performs roughly twice as many floating-point operations as the forward pass. Both factors make execution more compute-bound, spending a smaller fraction of time on the inter-GPU communication that MoK effectively overlaps or eliminates. Therefore, the largest relative gain appears in MXFP8 forward and the smallest in BF16 backward. We expect such gains to grow on future hardware; for example, Nvidia Vera Rubin \cite{nvidia2026rubingpu} raises FP8 compute throughput by 3.5$\times$ but scale-up network bandwidth by only 1.67$\times$ compared to GB300, making execution more communication-bound.

\begin{table}[t]
    \centering
    \vspace{-15pt}
    \caption{MoK sensitivity to EP degree and local token count. ``Tokens'' is local tokens per GPU before dispatch. ``Best'' is the lowest latency among the four baselines for the given workload, with the subscript identifying the best baseline (N for NCCL+PyTorch, D for DeepEP+PyTorch, H for HybridEP+Megatron). ``Speedup'' is ``Best'' divided by ``MoK''. Lower latency and higher speedup are better.}
    \label{tab:moe-sensitivity}
    \vspace{10pt}
    \small
    \setlength{\tabcolsep}{3pt}
    \begin{tabular}{@{}rlrrrrrrr@{}}
        \toprule
        & & & \multicolumn{2}{c}{Forward (ms)} & \multicolumn{2}{c}{Backward (ms)} & \multicolumn{2}{c}{Speedup} \\
        \cmidrule(lr){4-5}\cmidrule(lr){6-7}\cmidrule(l){8-9}
        EP & MoE Shape & Tokens & MoK & Best & MoK & Best & Fwd & Bwd \\
        \midrule
        \multirow{12}{*}[-6pt]{16} & \multirow{3}{*}{Kimi K2.7 Code} & 1{,}024 & 0.901 & 2.334$_{\mathrm{N}}$ & 1.786 & 2.413$_{\mathrm{N}}$ & \textbf{2.59$\times$} & \textbf{1.35$\times$} \\
        & & 2{,}048 & 1.332 & 2.581$_{\mathrm{N}}$ & 2.425 & 3.329$_{\mathrm{N}}$ & \textbf{1.94$\times$} & \textbf{1.37$\times$} \\
        & & 4{,}096 & 2.555 & 3.308$_{\mathrm{H}}$ & 4.320 & 5.053$_{\mathrm{H}}$ & \textbf{1.29$\times$} & \textbf{1.17$\times$} \\
        \cmidrule(l){2-9}
        & \multirow{3}{*}{GLM-5.2} & 1{,}024 & 0.849 & 2.321$_{\mathrm{N}}$ & 1.284 & 2.070$_{\mathrm{N}}$ & \textbf{2.73$\times$} & \textbf{1.61$\times$} \\
        & & 2{,}048 & 1.180 & 2.367$_{\mathrm{N}}$ & 1.965 & 2.762$_{\mathrm{N}}$ & \textbf{2.01$\times$} & \textbf{1.41$\times$} \\
        & & 4{,}096 & 2.051 & 2.913$_{\mathrm{H}}$ & 3.514 & 4.147$_{\mathrm{H}}$ & \textbf{1.42$\times$} & \textbf{1.18$\times$} \\
        \cmidrule(l){2-9}
        & \multirow{3}{*}{Qwen3.5-397B-A17B} & 1{,}024 & 0.857 & 2.316$_{\mathrm{N}}$ & 1.440 & 1.915$_{\mathrm{N}}$ & \textbf{2.70$\times$} & \textbf{1.33$\times$} \\
        & & 2{,}048 & 0.901 & 2.307$_{\mathrm{N}}$ & 1.296 & 2.129$_{\mathrm{N}}$ & \textbf{2.56$\times$} & \textbf{1.64$\times$} \\
        & & 4{,}096 & 1.705 & 2.777$_{\mathrm{N}}$ & 2.180 & 3.005$_{\mathrm{H}}$ & \textbf{1.63$\times$} & \textbf{1.38$\times$} \\
        \cmidrule(l){2-9}
        & \multirow{3}{*}{DeepSeek-V4-Pro} & 1{,}024 & 0.992 & 2.319$_{\mathrm{N}}$ & 2.321 & 2.571$_{\mathrm{N}}$ & \textbf{2.34$\times$} & \textbf{1.11$\times$} \\
        & & 2{,}048 & 1.560 & 2.612$_{\mathrm{N}}$ & 3.139 & 3.343$_{\mathrm{N}}$ & \textbf{1.67$\times$} & \textbf{1.06$\times$} \\
        & & 4{,}096 & 2.762 & 3.555$_{\mathrm{H}}$ & 5.157 & 5.462$_{\mathrm{D}}$ & \textbf{1.29$\times$} & \textbf{1.06$\times$} \\
        \midrule
        \multirow{12}{*}[-6pt]{64} & \multirow{3}{*}{Kimi K2.7 Code} & 1{,}024 & 1.016 & 2.600$_{\mathrm{N}}$ & 1.555 & 2.295$_{\mathrm{H}}$ & \textbf{2.56$\times$} & \textbf{1.48$\times$} \\
        & & 2{,}048 & 1.293 & 2.570$_{\mathrm{H}}$ & 2.045 & 2.636$_{\mathrm{H}}$ & \textbf{1.99$\times$} & \textbf{1.29$\times$} \\
        & & 4{,}096 & 2.524 & 4.002$_{\mathrm{H}}$ & 3.871 & 4.709$_{\mathrm{H}}$ & \textbf{1.59$\times$} & \textbf{1.22$\times$} \\
        \cmidrule(l){2-9}
        & \multirow{3}{*}{GLM-5.2} & 1{,}024 & 0.989 & 2.706$_{\mathrm{H}}$ & 1.612 & 2.287$_{\mathrm{H}}$ & \textbf{2.74$\times$} & \textbf{1.42$\times$} \\
        & & 2{,}048 & 1.142 & 2.471$_{\mathrm{H}}$ & 1.737 & 2.414$_{\mathrm{H}}$ & \textbf{2.16$\times$} & \textbf{1.39$\times$} \\
        & & 4{,}096 & 2.171 & 3.354$_{\mathrm{H}}$ & 3.294 & 4.147$_{\mathrm{H}}$ & \textbf{1.54$\times$} & \textbf{1.26$\times$} \\
        \cmidrule(l){2-9}
        & \multirow{3}{*}{Qwen3.5-397B-A17B} & 1{,}024 & 0.995 & 2.567$_{\mathrm{N}}$ & 1.442 & 2.348$_{\mathrm{H}}$ & \textbf{2.58$\times$} & \textbf{1.63$\times$} \\
        & & 2{,}048 & 1.068 & 2.527$_{\mathrm{H}}$ & 1.273 & 2.267$_{\mathrm{H}}$ & \textbf{2.37$\times$} & \textbf{1.78$\times$} \\
        & & 4{,}096 & 1.830 & 2.978$_{\mathrm{H}}$ & 2.137 & 2.959$_{\mathrm{H}}$ & \textbf{1.63$\times$} & \textbf{1.38$\times$} \\
        \cmidrule(l){2-9}
        & \multirow{3}{*}{DeepSeek-V4-Pro} & 1{,}024 & 0.981 & 2.670$_{\mathrm{N}}$ & 1.483 & 2.470$_{\mathrm{H}}$ & \textbf{2.72$\times$} & \textbf{1.67$\times$} \\
        & & 2{,}048 & 1.392 & 2.564$_{\mathrm{H}}$ & 2.321 & 2.839$_{\mathrm{H}}$ & \textbf{1.84$\times$} & \textbf{1.22$\times$} \\
        & & 4{,}096 & 2.594 & 3.796$_{\mathrm{N}}$ & 4.346 & 5.112$_{\mathrm{H}}$ & \textbf{1.46$\times$} & \textbf{1.18$\times$} \\
        \bottomrule
    \end{tabular}
    \vspace{-5pt}
\end{table}

\paragraph{Sensitivity to EP Degree and Local Token Count.} Table~\ref{tab:moe-sensitivity} reports MXFP8 latency and speedup for EP degrees 16 and 64, and for 1{,}024, 2{,}048, and 4{,}096 tokens per GPU. MoK achieves larger speedups at smaller local token counts, where communication and scheduling overheads account for a larger fraction of execution time. This regime is important in large-scale training, because the global batch size cannot be increased indefinitely without diminishing optimization returns. Scaling to more GPUs must therefore reduce the tokens processed per GPU~\cite{mccandlish2018empirical,shallue2019measuring,kaplan2020scaling}, making MoK increasingly effective as deployments scale.

\subsection{End-to-End Benchmarks}
\label{sec:End-to-end-benchmarks}

We also evaluate MoK in our internal production training stack. We compare against the previous version of the stack, which overlapped DeepEP-based MoE communication with our open-sourced MXFP8 expert FFN kernels \cite{sul2025mxfp8kernels}. We refer to this version as \textit{DeepEP-based}. Varying only the MoE implementation, we measure end-to-end training throughput, in tokens per second per GPU, across multiple GB300 NVL72 racks.

\begin{table}[!ht]
    \centering
    \caption{End-to-end production training stack throughput on 512 GB300 GPUs, comparing the DeepEP-based backend with MoK. Both runs use an EP degree of 32, MXFP8 routed experts, and a BF16 shared expert. We report the mean per-GPU throughput. Higher is better.}
    \label{e2e-throughput}
    \vspace{10pt}
    \small
    \begin{tabular}{@{}rcc@{}}
        \toprule
        & DeepEP-based & MoK \\
        \midrule
        Tokens / second / GPU & 761.0 & 1,070.2 \\
        \bottomrule
    \end{tabular}
\end{table}

As shown in Table~\ref{e2e-throughput}, replacing the DeepEP-based implementation with MoK increased end-to-end training throughput by 41\%. In practice, our production training runs use far more GPUs and fewer tokens per GPU than this experiment, which further increases MoK's gain, as explained in Section~\ref{sec:moe-layer-benchmarks}.

\section{Conclusion}
\label{sec:conclusion}

As AI accelerators consolidate into large scale-up domains, the design choices behind existing distributed MoE systems become increasingly suboptimal. We studied MoE training on NVL72 and identified three insights that unlock its performance: choosing the communication direction per operator, restructuring the computation-communication overlap around a tunable granularity, and eliminating CPU-GPU synchronization through device-side ring buffering. Mixture-of-Kittens (MoK) embodies these insights in a single deterministic megakernel, fusing MoE computation and communication with production-critical features such as hardware-accelerated MXFP8 execution, RDMA overlap for FSDP, and fused router weight gradient computation. MoK substantially outperforms the strongest publicly available baselines across the shapes of widely used open-weight models, and delivers end-to-end throughput gains in production-scale LLM training.

\section*{Acknowledgements}
We are grateful to Cursor for making this work possible. We thank
Vivien Cheng,
Dan Fu,
Hermann Kumbong,
Jerry Liu,
Chen Lu,
Sasha Rush,
Nathan Wang, and
Less Wright
for helpful feedback and discussions during this work.

\bibliographystyle{references}
\bibliography{references}

\appendix
\clearpage
\section*{Appendix}

\section{Scheduling Overhead}
\label{app:scheduling-overhead}

This section compares the scheduling overhead of MoK and Comet~\cite{zhang2025comet}. In MoK, scheduling involves the all-gather of the local top-$k$ routes followed by the dispatch schedule construction described in Algorithm~\ref{alg:dispatch-schedule-construction}. In Comet, scheduling involves the all-gather of the local top-$k$ routes followed by the per-expert token count and scatter index computation, the source-rank sort of the routes, and the host-side GEMM tile scheduling.

We measure BF16 forward passes at EP degree 4 with the workloads, inputs, and local token counts used in Section~\ref{sec:moe-layer-benchmarks}, timing 100 iterations after 500 warmup iterations and reporting the median slowest-rank latency. Absolute times are not comparable because MoK requires Blackwell GPUs and Comet supports GPUs only up to Hopper. Thus, we compare each system's relative scheduling overhead. Because the forward pass is faster on MoK, this comparison is biased \textit{against} MoK and favors Comet. Table~\ref{tab:scheduling-overhead} reports the results.

\begin{table}[H]
    \vspace{-5pt}
    \centering
    \caption{Scheduling overhead of MoK and Comet for BF16 forward passes at EP degree 4. ``Tokens'' is local tokens per GPU before dispatch. ``Sched.'' is the time spent building the schedule before dispatch begins, ``Total'' is ``Sched.'' plus the forward latency, both in ms, and ``Share'' is ``Sched.'' divided by ``Total''.}
    \label{tab:scheduling-overhead}
    \vspace{10pt}
    \small
    \setlength{\tabcolsep}{4pt}
    \begin{tabular}{@{}lrrrrrrr@{}}
        \toprule
        & & \multicolumn{3}{c}{MoK (GB300)} & \multicolumn{3}{c}{Comet (H100)} \\
        \cmidrule(lr){3-5}\cmidrule(l){6-8}
        MoE Shape & Tokens & Sched. & Total & Share & Sched. & Total & Share \\
        \midrule
        \multirow{3}{*}{Kimi K2.7 Code} & 1{,}024 & 0.068 & 2.250 & \textbf{3.04\%} & 0.331 & 4.892 & \textbf{6.76\%} \\
        & 2{,}048 & 0.093 & 2.571 & \textbf{3.60\%} & 0.389 & 5.640 & \textbf{6.89\%} \\
        & 4{,}096 & 0.133 & 4.507 & \textbf{2.96\%} & 0.515 & 10.537 & \textbf{4.89\%} \\
        \midrule
        \multirow{3}{*}{GLM-5.2} & 1{,}024 & 0.064 & 1.411 & \textbf{4.50\%} & 0.294 & 3.268 & \textbf{9.00\%} \\
        & 2{,}048 & 0.077 & 2.050 & \textbf{3.74\%} & 0.349 & 4.774 & \textbf{7.30\%} \\
        & 4{,}096 & 0.107 & 3.360 & \textbf{3.20\%} & 0.439 & 7.751 & \textbf{5.67\%} \\
        \midrule
        \multirow{3}{*}{Qwen3.5-397B-A17B} & 1{,}024 & 0.076 & 1.228 & \textbf{6.20\%} & 0.321 & 2.630 & \textbf{12.22\%} \\
        & 2{,}048 & 0.101 & 1.316 & \textbf{7.65\%} & 0.387 & 2.970 & \textbf{13.03\%} \\
        & 4{,}096 & 0.156 & 2.342 & \textbf{6.66\%} & 0.473 & 5.165 & \textbf{9.16\%} \\
        \midrule
        \multirow{3}{*}{DeepSeek-V4-Pro} & 1{,}024 & 0.068 & 3.195 & \textbf{2.14\%} & 0.325 & 6.878 & \textbf{4.72\%} \\
        & 2{,}048 & 0.087 & 3.468 & \textbf{2.50\%} & 0.361 & 7.614 & \textbf{4.75\%} \\
        & 4{,}096 & 0.115 & 5.201 & \textbf{2.22\%} & 0.504 & 11.927 & \textbf{4.23\%} \\
        \bottomrule
    \end{tabular}
\end{table}

\end{document}